%% file: main.tex
\documentclass[letterpaper]{article}
\usepackage{aaai24}
\usepackage{times}
\usepackage{helvet}
\usepackage{courier}
\usepackage[hyphens]{url}
\usepackage{graphicx}
\usepackage{natbib}
\usepackage{caption}
\usepackage{algorithm}
\usepackage{algorithmic}

\usepackage{graphicx}
\usepackage{booktabs}
\usepackage{multirow}
\usepackage{graphicx}
\usepackage{amsmath}
\usepackage{amsfonts}
\usepackage{mathrsfs}
\usepackage[capitalize,noabbrev]{cleveref}
\usepackage{tabularx}
\usepackage{multirow}
\usepackage{caption}
\usepackage{subcaption}
\usepackage{adjustbox}
\usepackage{makecell}
\usepackage{threeparttable}
\usepackage{algorithm}
\usepackage{algorithmic}
\usepackage{array}
\usepackage{xspace}
\usepackage{verbatim}
\usepackage{amsmath}
\usepackage{nicefrac}
\usepackage{comment}
\usepackage[table,xcdraw]{xcolor} 
\usepackage{dsfont}
\usepackage{xr}
\usepackage{subfiles}
\usepackage{enumitem}
\usepackage{mdframed}
\usepackage{tcolorbox}
\usepackage[T1]{fontenc}
\usepackage{float}

\usepackage{newfloat}
\usepackage{listings}
\DeclareCaptionStyle{ruled}{labelfont=normalfont,labelsep=colon,strut=off}
\floatstyle{ruled}
\newfloat{listing}{tb}{lst}{}
\floatname{listing}{Listing}

\title{Communicating Unexpectedness for Out-of-Distribution \\Multi-Agent Reinforcement Learning}
\author{
    Min Whoo Lee, Kibeom Kim, Soo Wung Shin, Minsu Lee, Byoung-Tak Zhang
}
\affiliations{
    Biointelligence Laboratory, Seoul National University\\
    1, Gwanak-ro, Gwanak-gu\\
    Seoul 08826 Republic of Korea\\
    \{mwlee,kbkim\}@bi.snu.ac.kr, ps4southwest@gmail.com, \{mslee,btzhang\}@bi.snu.ac.kr
}

\begin{document}

\maketitle

\begin{abstract}
Applying multi-agent reinforcement learning methods to realistic settings is challenging as it may require the agents to quickly adapt to unexpected situations that are rarely or never encountered in training.
Recent methods for generalization to such out-of-distribution settings are limited to more specific, restricted instances of distribution shifts.
To tackle adaptation to distribution shifts, we propose Unexpected Encoding Scheme, a novel decentralized multi-agent reinforcement learning algorithm where agents communicate ``unexpectedness,'' the aspects of the environment that are surprising.
In addition to a message yielded by the original reward-driven communication, each agent predicts the next observation based on previous experience, measures the discrepancy between the prediction and the actually encountered observation, and encodes this discrepancy as a message.
Experiments on multi-robot warehouse environment support that our proposed method adapts robustly to dynamically changing training environments as well as out-of-distribution environment.
\end{abstract}

\section{Introduction}

\begin{figure}[t]
    \centering
    \includegraphics[width=0.95\columnwidth]{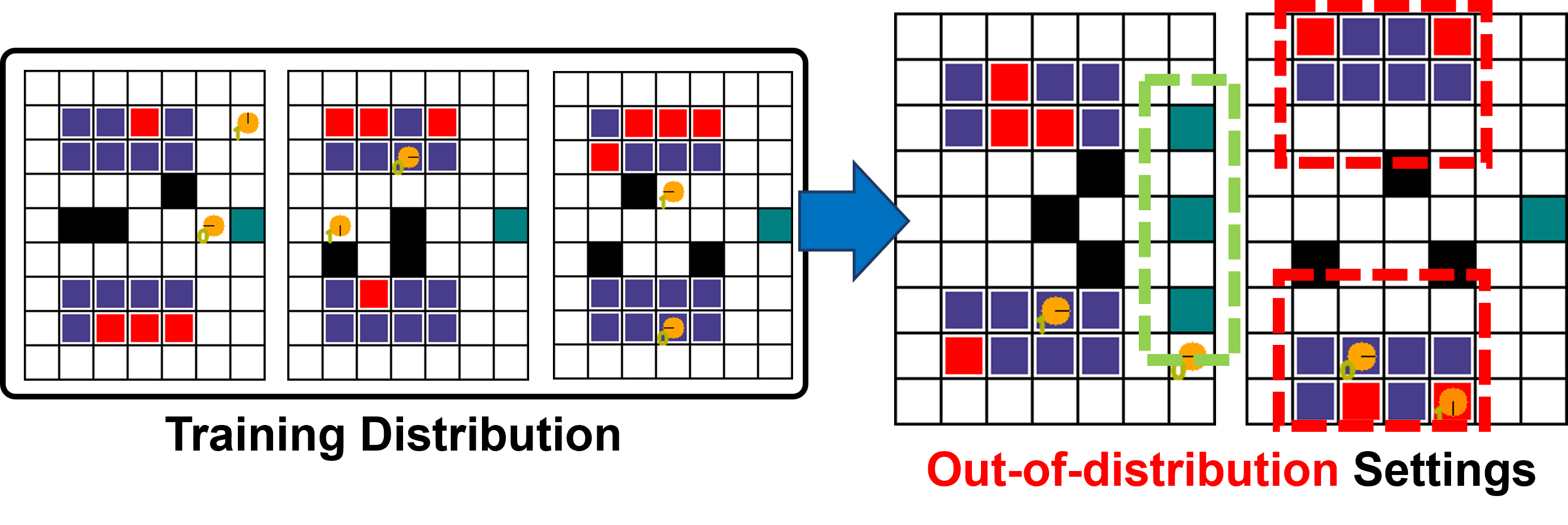}
    \caption{Conceptual diagram of the problem description.
    The green dotted box indicates the Goal-Shift setting, and the red dotted box indicates the Shelf-Shift setting.
    }
    \label{fig:problem}
\end{figure}

Recent development of multi-agent reinforcement learning (MARL) \cite{zhang2021multi} has shown promises in domains such as games \cite{schrittwieser2020mastering_muzero,vinyals2019grandmaster,berner2019dota}, unmanned aerial vehicles \cite{zhou2021multi_uav}, and smart grids \cite{zhang2022multistep_smartgrid}.
Nonetheless, extension of current MARL methods to reality still calls for multiple challenges.
First, the actual environments are very likely different from the environments on which the agents are trained, as illustrated in Figure \ref{fig:problem}.
In reality, agents will have to operate robustly even in those situations that they have never encountered within the distribution of the training environments, in order to avoid inefficient or even unsafe behaviours.
Meanwhile, given that one agent manages to adapt to some unexpected situation, it is important that other agents learn from this experience even without making these same mistakes.
Second, many realistic settings are partially observable and decentralized.
An individual agent is often only able to observe the vicinity, and the learning difficulty is exacerbated by the presence of other agents, who may alter their behaviours and environmental state dynamically.

These challenges call for novel use of communication in MARL.
In other words, while adopting RL to enable online adaptation of agents during unexpected, out-of-distribution situations in deployment, we also seek an effective inter-agent communication. 
Communication is a unique feature of multi-agent systems that allows agents to share information that may aid cooperative or distributed task-solving \cite{shoham2008multiagent}.
An appropriate use of agent-to-agent communication may resolve the aforementioned issues by sharing (1) environmental state information that is invisible to other agents, and (2) experience of unexpected situations that are not encountered during training.
Thus, our research objective is to develop an MARL architecture that effectively uses communication to let the agents adapt well in out-of-distribution downstream tasks.
Since agents will not have access to other agents' observations or actions in deployment, our method is designed as decentralized training, allowing the agents to adapt.

\begin{figure*}[t]
    \centering
      \captionsetup[subfigure]{oneside,margin={0cm,0cm}}
      \subfloat[][Overall architecture]{
        \includegraphics[width=0.44\linewidth]{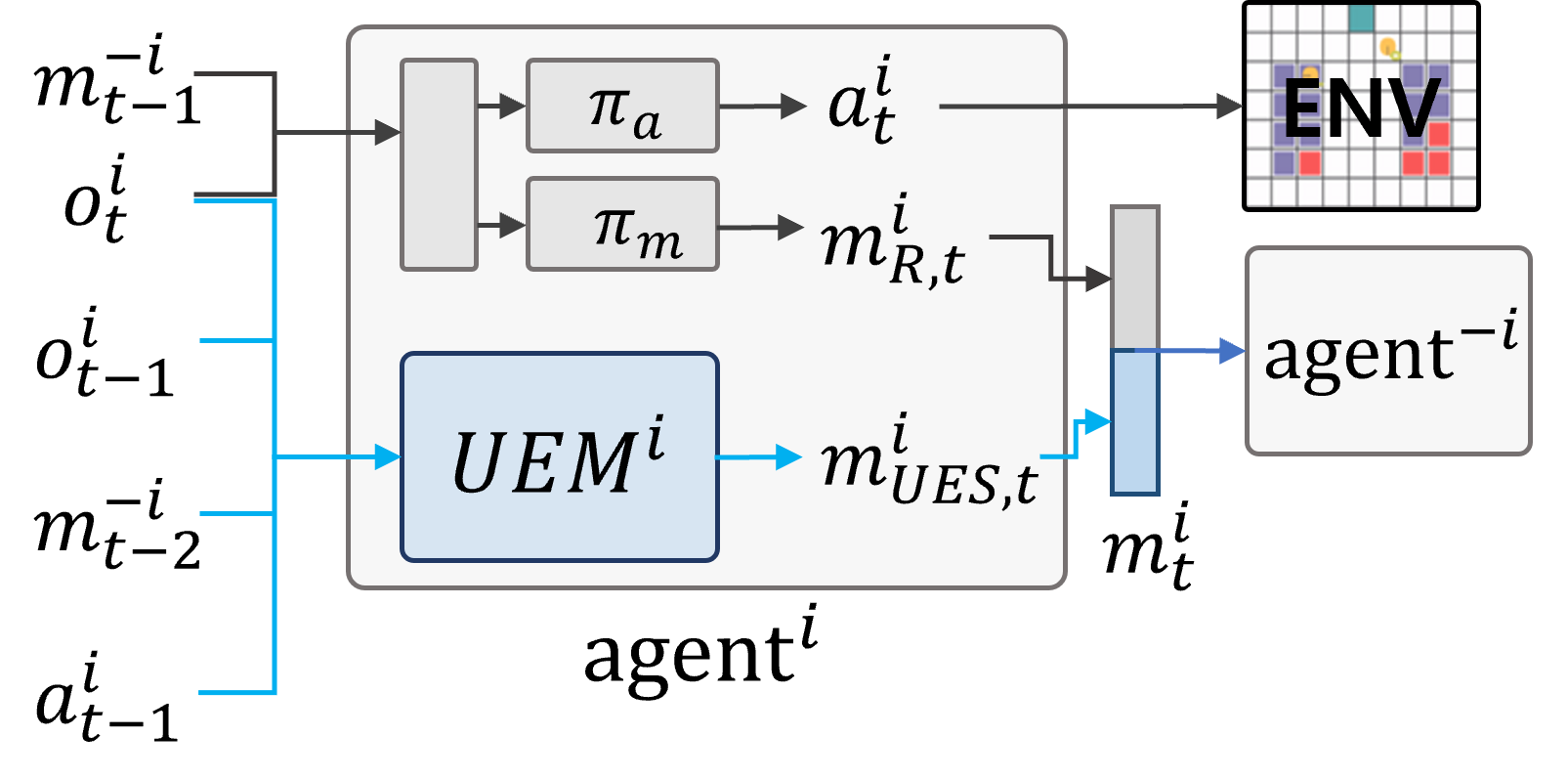}
        \label{fig:arch}
      }
      \subfloat[][Architecture of Unexpectedness Encoding Module (UEM)]{
        \includegraphics[width=0.45\linewidth]{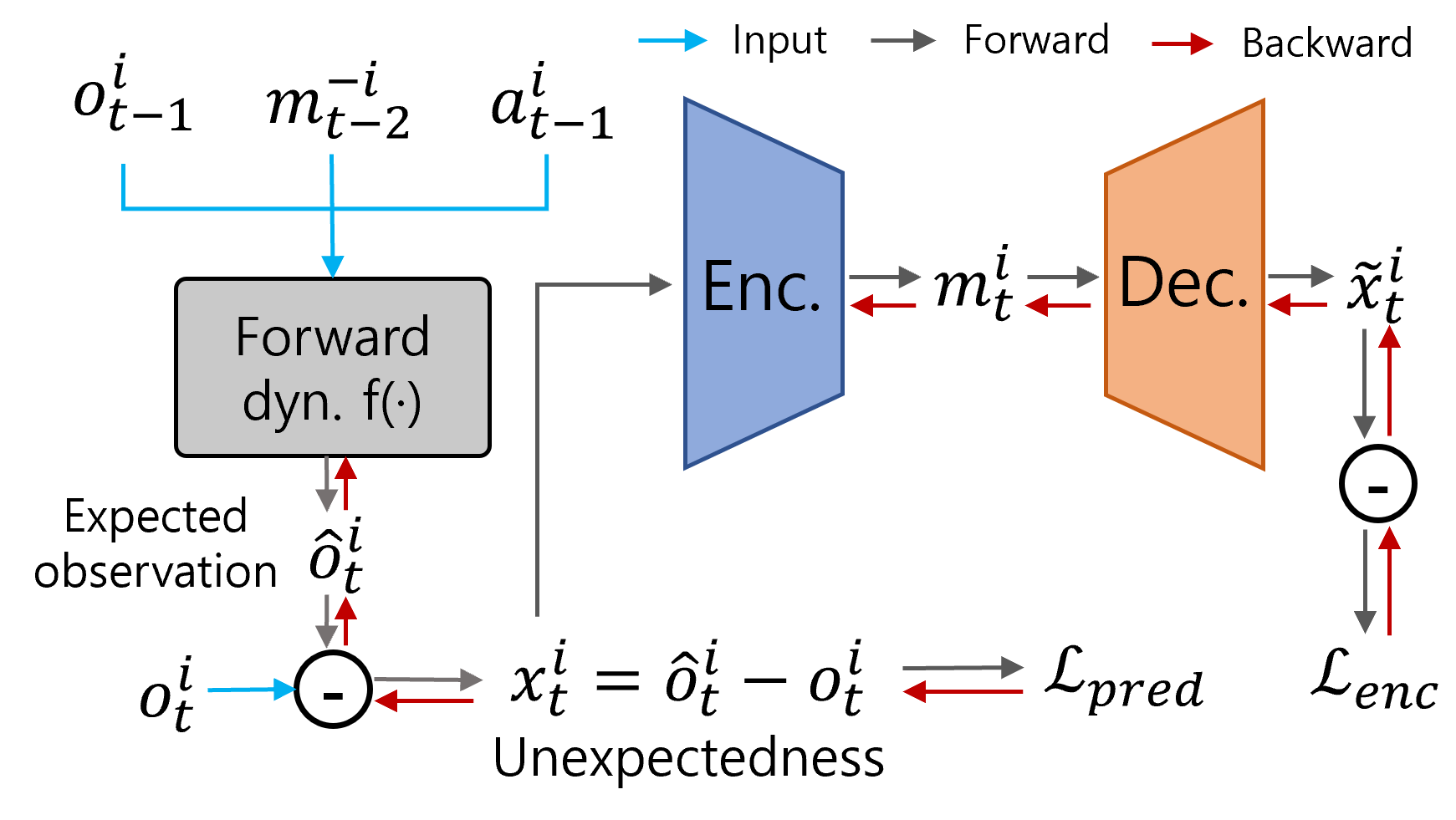}
        \label{fig:UEM}
      }
      \\
      \caption{Overview of the Unexpectedness Encoding Scheme with Reward (UES+R).
      }
      \label{fig:overall}
\end{figure*}

Recently, research has been extensively conducted on end-to-end learning of communication protocols in multi-agent systems \cite{zhu2022survey_commmarl,brandizzi2023towards_emcomsurvey}.
While vast amount of work has been done on investigating non-stationarity issue of multi-agent system that arises from evolving agent behaviors \cite{papoudakis2019dealing_nonstat,hernandez2017survey}, studies on generalizing multi-agent system to out-of-distribution environments are relatively scarce.
Particularly, several studies 
\cite{liu2021coach_copa,shao2022self_sog} demonstrate the capability of generalizing to unseen number of agents and environmental objects.
Nonetheless, these works assume that all environmental factors can be identified by objects and constrain the observation to a strict format.
It seems reasonable, on the other hand, that sudden environmental changes may occur in a more general manner that may not be restricted to identification of objects.
A recent work \cite{abu2021promoting_confusion} that is most closely related to ours introduces the notion of ``confusion'' to communicate unexpectedness of the environment.
However, the definition of confusion relies on immediate one-step reward, which is susceptible to noises in rewards and lacks consideration of long-term value.

We propose Unexpectedness Encoding Scheme with Reward (\textbf{UES+R}), a novel agent-to-agent communication scheme in decentralized MARL setting that enhances robustness to distribution shifts.
On top of the base communication scheme that is trained to maximize reward, we additionally devise Unexpectedness Encoding Scheme.
The key idea is to predict the next observation via forward dynamics, and measure the difference between the prediction and the actual next observation.
Such discrepancy vector is formulated as the \textit{unexpectedness}, and an autoencoder is used to encode this vector as \textit{unexpectedness encoding}.
This is fused with reward-driven message, in order to ensure task-relevant information to be shared among agents.
Experiments on the multi-agent warehouse environments support that UES+R not only boosts training performance in dynamically changing settings, but also promotes robust adaptation in environments with distribution shift.
Notably, performance of this decentralized training method is on par with a centralized training method that has access to all agents' observations.

\section{Unexpectedness Encoding for Communicating in Out-of-Distribution Environments}

\subsection{Overview}

We consider Decentralized Partially Observable Markov Decision Process (Dec-POMDP) \cite{oliehoek2008optimal}, a variant of Markov game \cite{littman1994markov} that accounts for partial observability.
At time step $t$, agent $i$ receives local observation $o_t^i$ from observation space $\mathcal{O}^i$ and messages from other agents $m_{t-1}^{-i}$, and chooses an action $a_t^i$ from action space $\mathcal{A}^i$.
In such partially observable setting, the local observations $o_t^i$ do not fully reflect the environment's true state $s_t \in \mathcal{S}$.
This necessitates an effective usage of messages to share individual agents' knowledge of surroundings and coordinate their behaviors.
The detailed preliminaries including Markov game and Dec-POMDP are covered in Appendix \ref{app:prelim}.

On top of using message to communicate task-relevant information among the agents, we also intend to use message as a feature that expresses how much and in what sense the environment differs from anticipation.
Consequently, if an out-of-distribution observation is encountered, the agent may encode this as message $m_t^i$ and communicate to the other agents, thus informing them about the surprising change in the environment.
In such case, the entire multi-agent system may be able to utilize the received message to learn about and prepare against such distribution shift situations.
To implement the aforementioned purpose, we propose a novel communication scheme called Unexpectedness Encoding Scheme with Reward (UES+R).
The architecture and data flow for each agent $i\in\mathcal{N}$ are outlined in Figure~\ref{fig:overall}.

\subsection{Unexpectedness Encoding Scheme}
\label{sec:ues}

First of all, we describe Unexpectedness Encoding Scheme (UES), the main addition of our proposed architecture.
This is illustrated in Figure~\ref{fig:UEM}.
To formulate the difference between anticipated and actual situations, we utilize a forward dynamics module $f(\cdot)$ that carries out the anticipation.
UES introduces an Unexpectedness Encoding Module (UEM) that uses the information from time step $(t-1)$ in order to predict the observation at time $t$ in hindsight.
To be specific, at time step $t$, UEM receives the agent $i$'s observation $o_{t-1}^i$, action $a_{t-1}^i$, and other agents' messages $m_{t-2}^{-i}$ at previous time steps.
Based on these inputs, the module performs forward dynamics $f(\cdot)$ and calculates the agent's prediction of the next observation $o_t^i$ in hindsight, denoted as $\hat{o}_t^i := f(o_{t-1}^i, m_{t-2}^{-i}, a_{t-1}^i)$.
We remind that the messages chosen at time step $(t-2)$ are received by the agent at time step $(t-1)$, hence the usage of $m_{t-2}^{-i}$.
The discrepancy between the prediction $\hat{o}_t^i$ and the actual observation $o_t^i$ is referred to as \textit{unexpectedness}\footnote{To be precise, as described in Appendix \ref{app:impl}, linear projection is used on the observations, rather than directly calculating the difference between raw observations.}, denoted as $x_t^i = \hat{o}_t^i - o_t^i$.

Then, in order to communicate the unexpected information relevant to distribution shift, the unexpectedness vector $x_t^i$ is encoded to yield an \textit{unexpectedness encoding}. This is the message output $m_{UES,t}^i$ of the UEM.
The encoding is executed by an autoencoder, used for both dimension reduction and learning to encode the unlabeled data.
In summary, the encoding process is represented by Equation \ref{eq:ues}.
\begin{equation}
    m_{UES,t}^i = \text{Enc} \left( f(o_{t-1}^i, m_{t-2}^{-i}, a_{t-1}^i) - o_t^i \right)
    \label{eq:ues}
\end{equation}

The forward dynamics model $f(\cdot)$ is trained by backpropagating the prediction loss $\mathcal{L}_{pred}$ in Equation \ref{eq:pred_loss}, which is the $l^2$-norm of the unexpectedness $x_t^i$.
The autoencoder is trained with the loss $\mathcal{L}_{enc}$ in Equation \ref{eq:enc_loss}, which is backpropagated only up to the encoder.
$\tilde{x}_t^i$ denotes the reconstruction output of the autoencoder.
\begin{gather}
    \mathcal{L}_{pred} := \left\| \hat{o}_t^i - o_t^i \right\|_2 = \left\| f(o_{t-1}^i, m_{t-2}^{-i}, a_{t-1}^i) - o_t^i \right\|_2 \label{eq:pred_loss} \\
    \mathcal{L}_{enc} := \left\| \tilde{x}_t^i - x_t^i \right\|_2 = \left\| \text{Dec}(\text{Enc}(x_t^i)) - x_t^i \right\|_2 \label{eq:enc_loss}
\end{gather}

\subsection{Incorporating Extrinsic Reward}

Intuitively, the message $m_{UES,t}^i$ yielded by UES can be deemed by the receiving agents as information that is useful for promptly learning about the environmental changes.
Nonetheless, since UES is driven purely by observation prediction error and not by environmental rewards, the unexpectedness encoding may not focus on including information that is indeed relevant to the task.
For the agents to conduct the task, communication guided by reward is needed.

For this, we consider a simple setting for reward-driven communication\footnote{Such reward-driven communication is a common setting, an example of which is in \cite{jaques2019social} for actor-critic methods.}.
A message $m_{R,t}^i$  is assumed to be a binary  bit vector $[u_1, u_2, \cdots, u_K ]$ of length $K$, such that $u_k \in \{0, 1\}$ for $i\in \{1,2, \cdots, K\}$.
Just as how a reinforcement learning agent is trained to choose its action via an objective that reflects the environment reward, the agent is trained to choose each bit $u_k$ of the message via the same RL objective.
In other words, each message bit is treated as a separate action channel and reinforced according to reward likewise.

The objective is determined by the specific algorithm that is used to train the agents.
For instance, we conduct our experiments using Advantage Actor-Critic (A2C) \cite{mnih2016a3c} whose loss gradients are outlined below for agent $i$:
\begin{gather}
    \nabla_\theta \mathcal{L}(\theta) = \sum_{t=1}^T \nabla_\theta \log \pi(a_t,m_{R,t}^i | o_t ; \theta) (R_t - V(o_t ; \phi)), \label{eq:a2c_actor_loss} \\
    \nabla_\phi \mathcal{L}_v (\phi) = \sum_{t=1}^T \nabla_\phi (R_t - V(o_t ; \phi))^2, \label{eq:a2c_critic_loss}
\end{gather}
where $\theta$ is actor parameters, $\phi$ is value parameters, $V$ is an individual agent-wise value function, and $R_t = \sum_{t'=t}^T \gamma^{t'-t} r_t$ is sum of rewards.
For simplicity, we omitted the the superscript $i$, which indicates the agent index, from $o_t^i$, $a_t^i$, $\theta^i$, $\phi^i$, and $\pi^i$.

To combine the benefit of UES in adapting to environmental changes and the usefulness of reward-driven communication in fulfilling the task, we fuse the messages from both schemes by concatenating them.
We refer to this combined scheme as \textbf{UES+R}, highlighting this as our main proposed method.

\begin{figure}
    \centering
    \includegraphics[width=0.9\columnwidth]{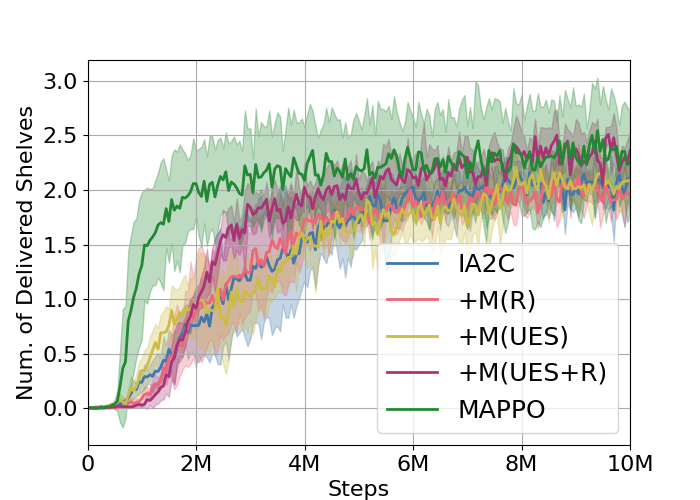}
    \caption{Learning curves on training distribution.
    Mean and standard deviation across 5 runs are plotted.
    Note that \textbf{M(UES+R)} is our main method, and MAPPO is intended to indicate the upper bound of performance.
    }
    \label{fig:learning_curve}
\end{figure}

\section{Experiments}

\subsection{Benchmark Environment}
\label{sec:bm_env}

We adopt Multi-Robot Warehouse (RWARE) environment \cite{papoudakis2020benchmarking} as our benchmark for our algorithm.
This environment, illustrated by snapshots in Figure \ref{fig:problem}, simulates grid-based robotic warehouses in reality, composed of two agents (orange circles) and many shelves containing items (blue/red squares).
Briefly described, the agents must deliver the requested shelves (red squares) to the goal location (teal tile), and subsequently return the shelf back to the original place.
To make the problem dynamically changing, three impassable obstacles (black tiles) are randomly positioned every 1K steps.
One challenging factor of this environment is that a long sequence of correct actions must be conducted to receive a reward, which means that reward is highly sparse.
Additionally, each agent can only observe adjacent tiles, making RWARE a Dec-POMDP.
The detailed environment setting is described in Appendix \ref{app:benchmark}.

\subsection{Out-of-Distribution Settings for Few-Shot Transfer Learning}
\label{sec:ood_few_shot}

The environment described in Section~\ref{sec:bm_env} outlines the \textit{training distribution}.
Our objective is to investigate the robustness of our method and the baselines to out-of-distribution environments that are not encountered during training.
For this, the few-shot transfer learning capability of the methods are measured.
In detail, the agents are initially trained on the training distribution for 10M steps.
Subsequently, the trained agents are placed in previously unseen distributions of environments, and fine-tuning is conducted for 10 batches of episodes.
The performance is recorded as the cumulative average number of successfully delivered shelves per episode.

We emphasize that the unseen distributions are designed such that the environment varies in a manner that has never been encountered during training.
We devise two such distributions as shown in Figure~\ref{fig:problem}: (1) \textbf{Goal-Shift}, with more goal tiles than in training, and (2) \textbf{Shelf-Shift}, with shelves' positions shifted towards the walls.
The former can be considered a beneficial distribution shift that imitates a realistic situation where new outlets are constructed for better throughput.
Agents that are robust to such distribution shifts of the environment will quickly adapt to and exploit the newly added goal tiles.
In contrast, Shelf-Shift increases the difficulty of the task, since the shelves adjacent to the wall are blocked by the surrounding shelves and are highly difficult to carry out.

\begin{table}[]
 \caption{
 Performances of baselines and the proposed methods. 
 Mean and standard deviation of number of delivered shelves across 5 runs are recorded.
 MAPPO is a centralized training method, intended to show an upper bound performance.
 }
 \label{tab:performances}
    \centering
    \begin{tabular}{l|rrr}
    \toprule
    Algorithm  & Training & Goal-Shift & Shelf-Shift \\
    \midrule
    IA2C                 & 2.16$\pm$0.12 & 2.04$\pm$0.13 & 1.34$\pm$0.04 \\
    +M(R)                & 2.12$\pm$0.30  & 2.04$\pm$0.14 & 1.33$\pm$0.12 \\
    +M(UES)                & 2.21$\pm$0.20 & 2.04$\pm$0.25 &  1.25$\pm$0.09\\
    +\textbf{M(UES+R)}      & \textbf{2.51}$\pm$\textbf{0.07} & \textbf{2.29}$\pm$\textbf{0.07} & \textbf{1.45}$\pm$\textbf{0.07}\\
    \midrule
    MAPPO         &  \multirow{2}{*}{2.54$\pm$0.49} & \multirow{2}{*}{2.45$\pm$0.50} & \multirow{2}{*}{1.31$\pm$0.22} \\
    (upper bound) & & & \\
    \bottomrule
    \end{tabular}
\end{table}

\subsection{Baselines}

The experiments were conducted with the following algorithms: (1) Independent Advantage Actor-Critic (\textbf{IA2C}), where A2C \cite{mnih2016a3c} was used to train agents individually without any message-sharing; (2) \textbf{IA2C+M(R)}, an ablative baseline where messages were trained via environmental reward $r_{t+1}^i$; (3) \textbf{IA2C+M(UES)}, another ablative baseline where messages are constructed only via UES; (4) \textbf{IA2C+M(UES+R)}, the proposed main method that fuses messages from the UES and reward; and (5) Multi-Agent Proximal Policy Optimization (\textbf{MAPPO}) \cite{yu2021surprising}, a multi-agent version of Proximal Policy Optimization algorithm \cite{schulman2017ppo} that has shown one of the highest performances in this task \cite{papoudakis2020benchmarking}.
It should be noted that all methods conduct decentralized training except for MAPPO, which is a centralized training method that has access to all agents' observations during training.
Rather, MAPPO can be considered as an upper bound of the performance for decentralized training methods.
We also clarify that the length of combined messages from \textbf{IA2C+M(UES+R)} is made to match the length of message from IA2C+M(R) and that from IA2C+M(UES) for fair comparison.
Additional details of these baselines are explained in Appendix~\ref{app:impl}.

\subsection{Results}
\label{sec:results}

The training of all five methods were conducted on the training distribution with the aforementioned methods for 10M steps.
Five repeated trials were conducted with different random seeds, and the learning curves and the highest performances are recorded in Figure \ref{fig:learning_curve} and Table \ref{tab:performances}, respectively.
It is fascinating that in the training distribution, the communication methods IA2C+M(R) and IA2C+M(UES) do not significantly surpass IA2C that does not communicate, which is evident from comparison of the learning curves.
This indicates that each of the two schemes alone does not lead to messages that beneficially contributes to solving the given task.
However, the complementarity of the two message schemes is shown, as \textbf{IA2C+M(UES+R)} converges to a superior performance that eventually attains the performance of MAPPO, the centralized-training upper bound method.

Furthermore, we evaluated the robustness of the trained agents to distribution shifts as outlined in Section~\ref{sec:ood_few_shot}.
The post-adaptation performances of all methods were recorded in Table \ref{tab:performances}.
The results showed a trend similar to one in the training distribution.
In other words, merely using reward or UES individually led to performances on par with or even worse than IA2C without messages, but a simultaneous use of both schemes led to noticeable improvement in robustness.
It is noteworthy that in Shelf-Shift experiment, MAPPO, which we expected to be an upper bound, performed similarly to IA2C, a decentralized method.
In comparison, our proposed \textbf{IA2C+M(UES+R)} achieves the highest performance.
This indicates that simply sharing all agents' observations and conducting centralized training may not remedy the overfitting to the training distribution.
Rather, the key solution may be the communication scheme that discreetly selects the messages that specifically inform how the environment has changed -- addressed by UES -- and what aspect of this new environment is relevant to task-solving -- addressed by reward-driven message.
We believe the result supports that our method \textbf{IA2C+M(UES+R)} achieves both.

\section{Conclusion}
\label{sec:conclusion}

We proposed UES+R, a novel MARL scheme that communicates unexpectedness to adapt to environmental distribution shifts.
This scheme measures the difference between predicted observation and actual observation, and encodes this discrepancy as the message that is received by other agents.
This message is combined with message that is trained to maximize reward.
Experiments on multi-robot warehouse environment indicate that our proposed method leads to robust adaptation to dynamic, out-of-distribution environment.

Nonetheless, several limitations lie within this study.
For instance, we assume the setting that messages are broadcasted to all agents, which is infeasible when number of agents are high.
A method to compress or encode these into smaller representations may be needed, such as an attentional method in \cite{jiang2018learning}.
Also, interpreting the message contents in relation to actual environmental shift is an important topic for future work.

\section*{Acknowledgments}
The authors are thankful to Hyundo Lee, Won-Seok Choi, and the anonymous reviewers for their valuable feedback on the drafts of this paper.
This work was partly supported by the IITP (RS-2021-II212068-AIHub/10\%, 2021-0-01343-GSAI/10\%, 2022-0-00951-LBA/20\%, 2022-0-00953-PICA/25\%) and NRF (RS-2023-00274280/10\%, 2021R1A2C1010970/25\%) grant funded by the Korean government.

\bibliography{aaai24}

\appendix
\input{supplementary}

\end{document}

%% file: supplementary.tex
\section{Appendix}

\subsection{Preliminaries}
\label{app:prelim}

A common formulation of multi-agent reinforcement learning problem is Markov game \cite{littman1994markov}, which can be considered an extension of single-agent Markov Decision Process (MDP) to multi-agent setting.
Markov game is defined as a  tuple $ (\mathcal{N}, \mathcal{S}, \{\mathcal{A}^i\}_{i\in\mathcal{N}}, \mathcal{P}, \{\mathcal{R}^i\}_{i\in\mathcal{N}}, \gamma), $
where $\mathcal{N} = \{1,...,N\}$ is the set of agent indices with $N$ agents, $\mathcal{S}$ is state space, $\mathcal{A}^i$ is action space of agent $i$, and $\gamma$ is discount factor.
For convenience, we denote joint action space as $\mathcal{A} = \mathcal{A}^1 \times \cdots \times \mathcal{A}^N$.
Also, $\mathcal{P}: \mathcal{S} \times \mathcal{A} \rightarrow \Delta\mathcal{S}$ is transition probability, $\mathcal{R}^i: \mathcal{S} \times \mathcal{A} \rightarrow \mathbb{R}$ is reward function of agent $i$.
Every agent $i$ has access to state $s_t \in \mathcal{S}$, based on which the agent can choose action $a_t^i$ according to its policy $\pi^i: \mathcal{S} \rightarrow \Delta(\mathcal{A}^i)$.
Then, the agent receives the next state $s_{t+1} \sim \mathcal{P}(s_{t+1} | s_t, \mathbf{a}_t)$ and reward $r_{t+1}^i = \mathcal{R}^i(s_t,\mathbf{a}_t)$, where $\mathbf{a}_t = (a_t^1, ..., a_t^N)$ is joint action at time step $t$.

The aim of each agent is to select a policy that maximizes individual value function
\begin{equation}
	V^{\pi^i,\pi^{-i}}(s) = \mathbb{E}_\pi \left[\sum_{t'=t}^T \gamma^{t'-t} \mathcal{R}^i (S_{t'}, \mathbf{A}_{t'}) \middle\vert S_t = s_t \right],
\end{equation}
where $\pi(\mathbf{a}_t|s_t) = \prod_{i=1}^N \pi^i (a_t^i | s_t)$ is combination of all agents' policies, and $\pi^{-i}$ is combination of all agents' policies excluding the policy of agent $i$.
In Markov games, Nash Equilibrium is widely used as the notion of optimality and is defined as joint policy $(\pi^{1,*}, \cdots, \pi^{N,*})$ where, for each $i \in \mathcal{N}$,
\begin{equation}
\forall{s \in \mathcal{S}},\pi^i, ~~~ V^{\pi^{i,*},\pi^{-i,*}}(s) \geq V^{\pi^i,\pi^{-i,*}}(s).
\end{equation}
Intuitively, this means that every agent $i$ is discouraged from choosing some other policy $\pi^i \neq \pi^{i,*}$ given the fixed policies of other agents $\pi^{-i,*}$.

We consider Decentralized Partially Observable Markov Decision Process (Dec-POMDP), a variant of Markov game that accounts for partial observability \cite{oliehoek2008optimal}.
Its definition is similar to that of Markov game, with the addition of the set of agent observations $\{\mathcal{O}^i\}_{i\in\mathcal{N}}$ and observation functions $\Omega^i:\mathcal{S} \rightarrow \mathcal{O}^i$ that maps state   to individual agent's observation.
An agent $i$ has no access to the true global state $s_t$, and instead receives an observation $o_t^i := \Omega^i(s_t)$.

To enable agent-to-agent communication, we further allow a finite set of messages $\{\mathcal{M}^i\}_{i\in\mathcal{N}}$, from which each agent $i$ can choose $m_t^i \in \mathcal{M}^i$ to communicate to other agents.
In our problem setting, we assume that all agents' messages $m_t = (m_t^1, \cdots, m_t^N)$ at time step $t$ are broadcasted to all agents and are concatenated to the next observations of each agent $i\in \mathcal{N}$ as $\tilde{o}_{t+1}^i = [o_{t+1}^i, m_t]$ at the next time step.

\subsection{Implementation Details}
\label{app:impl}

\subsubsection{Common Details of All Baselines.}
This section describes the common settings that all baseline methods implemented on top of IA2C \cite{mnih2016a3c} follow.
Actor function of each agent is a neural network that contains a hidden linear layer of size 64 followed by a Gated-Recurrent Unit \cite{cho2014properties} of size 64.
The GRU is followed by another linear layer that outputs the logits for the actions, after which Softmax function is applied to yield the action probabilities.
Critic function contains a single linear layer of size 64.
Critic function learns to predict the $n$-step action value, where $n=5$.
The input to both the actor function and the critic function contains the current state and the messages of other agents.

Batch size of 10 is used.
Adam optimizer \cite{kingma2014adam} is used with $\alpha = 0.99$ and $\epsilon = 10^{-5}$.
Soft updates are applied to target critic networks with $\tau = 0.01$.
Rectified linear unit (ReLU) nonlinearity is applied to the output of every MLP or GRU.

\subsubsection{IA2C.}
The experiments with IA2C can be considered a control group that does not communicate any message.
Learning rate of 0.0005 is used for both the actor and the critic.
The objective function contains an entropy term with coefficient of 0.01.

\subsubsection{IA2C+M(R).}
This baseline serves as a naive decentralized method where the agents' communications are only driven by environmental rewards.
Learning rate of 0.0005 is used for both the actor and the critic.
The objective function contains an entropy term with coefficient of 0.01.
Message length of 10 is used, and each message bit is discretized to values of 0 or 1.
Each message bit is treated as a separate action channel, also trained with A2C.

\subsubsection{IA2C+M(UES).}
Learning rate of 0.001 is used for both the actor and the critic.
Message length of 10 is used, where each message element is a continuous value between 0 and 1.
The objective function contains an entropy term with coefficient of 0.01.

To conduct dimensionality reduction on the observation, a random linear projection $g(\cdot)$ was applied on both the previous observation $o_{t-1}^i$ and the current observation $o_t^i$.
The linear projection is implemented as a linear layer of output size 64, and ReLU is applied.

The forward dynamics module $f(\cdot)$ in Section~\ref{sec:ues} receives the local observation embedding $g(o_{t-1}^i$), the action $a_{t-1}^i$, and the messages of other agents $m_{t-2}^{-i}$.
This module is a 2-layered MLP, with both layers having an output size of 64.
Hence, to be precise, the unexpectedness $x_t^i$ in Equations \ref{eq:ues}, \ref{eq:pred_loss}, and \ref{eq:enc_loss} are calculated as $x_t^i = g(\hat{o}_t^i) - g(o_t^i)$.

The encoder is a linear layer with output size of 10, which is the length of the message, and the decoder is a linear layer with output size of 64 for reconstructing $x_t^i$.

\subsubsection{IA2C+M(UES+R).}
The UEM structure is identical to the one used in IA2C+M(UES), except for the message size.
The message yielded via  reward scheme is of length 5, and the message yielded via UES is of length 5, which leads to a concatenated message length of 10.
This matches the message length of other baselines, for fair comparison.

Learning rate of 0.0005 is used for both the actor and the critic.
The objective function contains an entropy term with coefficient of 0.05.

\subsection{Benchmark Environment Details}
\label{app:benchmark}

The environment is an adapted version of Multi-Robot Warehouse benchmark proposed in \cite{papoudakis2020benchmarking}.
Refer to the snapshots in Figure \ref{fig:problem}.
The five possible actions for each agent are \{MoveForward, RotateLeft, RotateRight, Pickup/PutDown, NoOp\}.
Among the shelves (blue squares), four are requested (red squares) to be delivered to the goal location (teal tiles).
An agent must then reach the tile containing a requested shelf, pick up the shelf, and carry it all the way to the goal location to receive a reward of +0.5.
The agent must also return the shelf to its original location, which grants an additional +0.5 reward.
Each episode is 50 time steps long, and agent positions are randomized at the start of every episode.

Each tile can be occupied by at most one agent.
If two agents try to move into the same tile, one agent is arbitrarily chosen to successfully move into the tile, and the other remains stationary.

An agent can move to a tile with a shelf and pick it up using the ``Pickup/Putdown'' action.
If the agent is currently holding a shelf, it can put down the shelf via the same action.
However, to avoid the shelves from being placed in narrow spaces such as corridors and blocking other shelves from being carried (one tile can only contain at most one shelf), the shelf can only be placed on tiles where the shelves are spawned.
The remaining tiles where the shelves cannot be placed on are called ``highways''.

An agent can only observe the information of immediately surrounding tiles, within the $3\times 3$ square centered on the agent.
This information includes whether each tile is occupied by an agent, which direction that agent is facing, whether the tile contains a shelf, whether that shelf is requested, and whether the tile contains an impassable obstacle.
Also, the agent observes its own grid position, the direction it is facing, whether it is carrying a shelf or not, and whether the current tile is a ``highway''.

To encourage each agent to carry requested shelves on their own while not specifically hindering the other agent from carrying the shelves, the reward function is designed such that the agent that successfully carries the requested shelf to the goal or returns the shelf to the original place receives the full reward of +0.5, while the other agent receives +0.125 reward.
This modification was adopted to simplify the credit assignment problem of the given high-difficulty sparse-reward environment.

%% file: main.bbl
\begin{thebibliography}{24}
\providecommand{\natexlab}[1]{#1}

\bibitem[{Abu et~al.(2021)Abu, Gerstgrasser, Rosenschein, and
  Keren}]{abu2021promoting_confusion}
Abu, O.; Gerstgrasser, M.; Rosenschein, J.; and Keren, S. 2021.
\newblock Promoting Resilience in Multi-Agent Reinforcement Learning via
  Confusion-Based Communication.
\newblock \emph{arXiv preprint arXiv:2111.06614}.

\bibitem[{Berner et~al.(2019)Berner, Brockman, Chan, Cheung, D{\k{e}}biak,
  Dennison, Farhi, Fischer, Hashme, Hesse et~al.}]{berner2019dota}
Berner, C.; Brockman, G.; Chan, B.; Cheung, V.; D{\k{e}}biak, P.; Dennison, C.;
  Farhi, D.; Fischer, Q.; Hashme, S.; Hesse, C.; et~al. 2019.
\newblock Dota 2 with large scale deep reinforcement learning.
\newblock \emph{arXiv preprint arXiv:1912.06680}.

\bibitem[{Brandizzi(2023)}]{brandizzi2023towards_emcomsurvey}
Brandizzi, N. 2023.
\newblock Towards More Human-like AI Communication: A Review of Emergent
  Communication Research.
\newblock \emph{arXiv preprint arXiv:2308.02541}.

\bibitem[{Cho et~al.(2014)Cho, Van~Merri{\"e}nboer, Bahdanau, and
  Bengio}]{cho2014properties}
Cho, K.; Van~Merri{\"e}nboer, B.; Bahdanau, D.; and Bengio, Y. 2014.
\newblock On the properties of neural machine translation: Encoder-decoder
  approaches.
\newblock \emph{arXiv preprint arXiv:1409.1259}.

\bibitem[{Hernandez-Leal et~al.(2017)Hernandez-Leal, Kaisers, Baarslag, and
  de~Cote}]{hernandez2017survey}
Hernandez-Leal, P.; Kaisers, M.; Baarslag, T.; and de~Cote, E.~M. 2017.
\newblock A survey of learning in multiagent environments: Dealing with
  non-stationarity.
\newblock \emph{arXiv preprint arXiv:1707.09183}.

\bibitem[{Jaques et~al.(2019)Jaques, Lazaridou, Hughes, Gulcehre, Ortega,
  Strouse, Leibo, and De~Freitas}]{jaques2019social}
Jaques, N.; Lazaridou, A.; Hughes, E.; Gulcehre, C.; Ortega, P.; Strouse, D.;
  Leibo, J.~Z.; and De~Freitas, N. 2019.
\newblock Social influence as intrinsic motivation for multi-agent deep
  reinforcement learning.
\newblock In \emph{International Conference on Machine Learning}, 3040--3049.
  PMLR.

\bibitem[{Jiang and Lu(2018)}]{jiang2018learning}
Jiang, J.; and Lu, Z. 2018.
\newblock Learning attentional communication for multi-agent cooperation.
\newblock \emph{Advances in neural information processing systems}, 31.

\bibitem[{Kingma and Ba(2014)}]{kingma2014adam}
Kingma, D.~P.; and Ba, J. 2014.
\newblock Adam: A method for stochastic optimization.
\newblock \emph{arXiv preprint arXiv:1412.6980}.

\bibitem[{Littman(1994)}]{littman1994markov}
Littman, M.~L. 1994.
\newblock Markov games as a framework for multi-agent reinforcement learning.
\newblock In \emph{Machine learning proceedings 1994}, 157--163. Elsevier.

\bibitem[{Liu et~al.(2021)Liu, Liu, Stone, Garg, Zhu, and
  Anandkumar}]{liu2021coach_copa}
Liu, B.; Liu, Q.; Stone, P.; Garg, A.; Zhu, Y.; and Anandkumar, A. 2021.
\newblock Coach-player multi-agent reinforcement learning for dynamic team
  composition.
\newblock In \emph{International Conference on Machine Learning}, 6860--6870.
  PMLR.

\bibitem[{Mnih et~al.(2016)Mnih, Badia, Mirza, Graves, Lillicrap, Harley,
  Silver, and Kavukcuoglu}]{mnih2016a3c}
Mnih, V.; Badia, A.~P.; Mirza, M.; Graves, A.; Lillicrap, T.; Harley, T.;
  Silver, D.; and Kavukcuoglu, K. 2016.
\newblock Asynchronous methods for deep reinforcement learning.
\newblock In \emph{International conference on machine learning}, 1928--1937.

\bibitem[{Oliehoek, Spaan, and Vlassis(2008)}]{oliehoek2008optimal}
Oliehoek, F.~A.; Spaan, M.~T.; and Vlassis, N. 2008.
\newblock Optimal and approximate Q-value functions for decentralized POMDPs.
\newblock \emph{Journal of Artificial Intelligence Research}, 32: 289--353.

\bibitem[{Papoudakis et~al.(2019)Papoudakis, Christianos, Rahman, and
  Albrecht}]{papoudakis2019dealing_nonstat}
Papoudakis, G.; Christianos, F.; Rahman, A.; and Albrecht, S.~V. 2019.
\newblock Dealing with non-stationarity in multi-agent deep reinforcement
  learning.
\newblock \emph{arXiv preprint arXiv:1906.04737}.

\bibitem[{Papoudakis et~al.(2020)Papoudakis, Christianos, Sch{\"a}fer, and
  Albrecht}]{papoudakis2020benchmarking}
Papoudakis, G.; Christianos, F.; Sch{\"a}fer, L.; and Albrecht, S.~V. 2020.
\newblock Benchmarking multi-agent deep reinforcement learning algorithms in
  cooperative tasks.
\newblock \emph{arXiv preprint arXiv:2006.07869}.

\bibitem[{Schrittwieser et~al.(2020)Schrittwieser, Antonoglou, Hubert,
  Simonyan, Sifre, Schmitt, Guez, Lockhart, Hassabis, Graepel
  et~al.}]{schrittwieser2020mastering_muzero}
Schrittwieser, J.; Antonoglou, I.; Hubert, T.; Simonyan, K.; Sifre, L.;
  Schmitt, S.; Guez, A.; Lockhart, E.; Hassabis, D.; Graepel, T.; et~al. 2020.
\newblock Mastering atari, go, chess and shogi by planning with a learned
  model.
\newblock \emph{Nature}, 588(7839): 604--609.

\bibitem[{Schulman et~al.(2017)Schulman, Wolski, Dhariwal, Radford, and
  Klimov}]{schulman2017ppo}
Schulman, J.; Wolski, F.; Dhariwal, P.; Radford, A.; and Klimov, O. 2017.
\newblock Proximal policy optimization algorithms.
\newblock \emph{arXiv preprint arXiv:1707.06347}.

\bibitem[{Shao et~al.(2022)Shao, Lou, Zhang, Jiang, He, and
  Ji}]{shao2022self_sog}
Shao, J.; Lou, Z.; Zhang, H.; Jiang, Y.; He, S.; and Ji, X. 2022.
\newblock Self-Organized Group for Cooperative Multi-agent Reinforcement
  Learning.
\newblock \emph{Advances in Neural Information Processing Systems}, 35:
  5711--5723.

\bibitem[{Shoham and Leyton-Brown(2008)}]{shoham2008multiagent}
Shoham, Y.; and Leyton-Brown, K. 2008.
\newblock \emph{Multiagent systems: Algorithmic, game-theoretic, and logical
  foundations}.
\newblock Cambridge University Press.

\bibitem[{Vinyals et~al.(2019)Vinyals, Babuschkin, Czarnecki, Mathieu, Dudzik,
  Chung, Choi, Powell, Ewalds, Georgiev et~al.}]{vinyals2019grandmaster}
Vinyals, O.; Babuschkin, I.; Czarnecki, W.~M.; Mathieu, M.; Dudzik, A.; Chung,
  J.; Choi, D.~H.; Powell, R.; Ewalds, T.; Georgiev, P.; et~al. 2019.
\newblock Grandmaster level in StarCraft II using multi-agent reinforcement
  learning.
\newblock \emph{Nature}, 575(7782): 350--354.

\bibitem[{Yu et~al.(2021)Yu, Velu, Vinitsky, Wang, Bayen, and
  Wu}]{yu2021surprising}
Yu, C.; Velu, A.; Vinitsky, E.; Wang, Y.; Bayen, A.; and Wu, Y. 2021.
\newblock The Surprising Effectiveness of PPO in Cooperative, Multi-Agent
  Games.
\newblock \emph{arXiv preprint arXiv:2103.01955}.

\bibitem[{Zhang, Yang, and Ba{\c{s}}ar(2021)}]{zhang2021multi}
Zhang, K.; Yang, Z.; and Ba{\c{s}}ar, T. 2021.
\newblock Multi-agent reinforcement learning: A selective overview of theories
  and algorithms.
\newblock \emph{Handbook of Reinforcement Learning and Control}, 321--384.

\bibitem[{Zhang et~al.(2022)Zhang, Yang, An, Li, and
  Wu}]{zhang2022multistep_smartgrid}
Zhang, Y.; Yang, Q.; An, D.; Li, D.; and Wu, Z. 2022.
\newblock Multistep multiagent reinforcement learning for optimal energy
  schedule strategy of charging stations in smart grid.
\newblock \emph{IEEE Transactions on Cybernetics}.

\bibitem[{Zhou et~al.(2021)Zhou, Liu, Li, Xu, and Shen}]{zhou2021multi_uav}
Zhou, W.; Liu, Z.; Li, J.; Xu, X.; and Shen, L. 2021.
\newblock Multi-target tracking for unmanned aerial vehicle swarms using deep
  reinforcement learning.
\newblock \emph{Neurocomputing}, 466: 285--297.

\bibitem[{Zhu, Dastani, and Wang(2022)}]{zhu2022survey_commmarl}
Zhu, C.; Dastani, M.; and Wang, S. 2022.
\newblock A survey of multi-agent reinforcement learning with communication.
\newblock \emph{arXiv preprint arXiv:2203.08975}.

\end{thebibliography}
